\documentclass[twocolumn,aps,prb,showpacs,superscriptaddress,longbibliography,10pt]{revtex4-2} 
\usepackage[colorlinks=true,linkcolor=blue,citecolor=blue]{hyperref}
\usepackage{amsfonts}
\usepackage{subfigure}
\usepackage{amsmath}
\usepackage{mathrsfs}
\usepackage{amssymb}
\usepackage{amsbsy}
\usepackage{epsfig}
\usepackage{graphicx}
\usepackage{epstopdf}
\usepackage{mathdots}
\usepackage{color,xcolor}
\usepackage{cleveref}
\usepackage{float}
\usepackage{graphicx}
\usepackage{rotating}
\usepackage{physics}
\usepackage{nicefrac}
\usepackage{soul}
\sethlcolor{yellow}
\setstcolor{blue}
\setulcolor{blue}
\usepackage[colorlinks=true, linkcolor=blue]{hyperref}
\usepackage{cleveref}
\usepackage{soul}
\soulregister\ref7 
\soulregister\citealt7 
\usepackage{cancel}

\begin{document}

\title{Interplay of Quasiperiodic Criticality and the Non-Hermitian Skin Effect}

\author{Zhangyuan Chen}
\affiliation{Department of Physics, Jiangsu University, Zhenjiang, 212013, China}

\author{Xianqi Tong} \altaffiliation{xqtong@ujs.edu.cn}
\affiliation{Department of Physics, Jiangsu University, Zhenjiang, 212013, China}

\author{Xiaosen Yang} \altaffiliation{yangxs@ujs.edu.cn} 
\affiliation{Department of Physics, Jiangsu University, Zhenjiang, 212013, China}

\begin{abstract}
Quasiperiodic lattices can host critical eigenstates, whereas nonreciprocal hopping in non-Hermitian lattices can induce non-Hermitian skin effect. In this work, we investigate localization phenomena in a Hatano--Nelson model with quasiperiodically modulated hopping amplitudes, where nonreciprocity arises from unequal modulation strengths of the right and left hoppings. Using a non-unitary gauge transformation, we map the non-Hermitian system into a Hermitian quasiperiodic system and obtain an exact analytical expression for the Lyapunov exponent in the thermodynamic limit. Under periodic boundary conditions, inverse participation ratios and finite-size scaling analysis are used to identify the quasiperiodic critical regimes. The comparison shows that parameter regimes hosting quasiperiodic critical states under periodic boundary conditions can exhibit the non-Hermitian skin effect under open boundary conditions. Furthermore, the non-Hermitian skin effect associated with quasiperiodic critical regimes is also observed in representative long-range hopping models and multiband extensions. Our results provide an analytically controlled perspective on how quasiperiodicity, modulated nonreciprocity, and boundary conditions jointly shape the non-Hermitian skin effect in critical regimes.
\end{abstract}

\maketitle
\section{Introduction}
Non-Hermiticity introduces novel phenomena distinct from their Hermitian counterparts~\cite{Bender1998Real,Zeuner2015Observation,Leykam2017Edge,Lee2019Topological,Ashida2020NonHermitian,Zhang2020Correspondence,li2020critical,zhang2021observation,Li2022Topological,jiGeneralized2024, FuPRBbraiding, Li2025Exact,Zhang2025Yang-Lee,Wang2025Theory}, most notably the non-Hermitian skin effect (NHSE)~\cite{Okugawa2020Second-order, Yokomizo2021Scaling, Yiling2022Flexible, Gu2022TransientNHSE, Liang2022Dynamic, lin2023topological, Li2024DynamicNHSE, Yoshida2024Non-Hermitian, Ma2024Non-Hermitian, Lin2024Observation, Wang2025Nonlinear, Wang2025Tunable, yang2026kagome, Wang2026One-Dimensional}. This boundary accumulation of bulk eigenstates profoundly modifies the conventional bulk-boundary correspondence and has stimulated extensive studies on boundary-dependent localization in non-Hermitian systems~\cite{Yao2018Edge,Kunst_Biorthogonal2018,Yokomizo_Non-Bloch2019,Okuma2020Topological,Bergholtz2021Exceptional,zhang2022universal,zhou2023observation,Wang2024Amoeba}. The localization properties of non-Hermitian systems can depend strongly on boundary conditions~\cite{Yao2018Edge,Yokomizo_Non-Bloch2019,Okuma2020Topological,zhang2022universal}. Eigenstates under open boundary conditions (OBC) and periodic boundary conditions (PBC) may exhibit qualitatively different behaviors~\cite{Lee2016Anomalous,Song2019Chiral,Borgnia2020Boundary,Nakai2024Topological,Padhi2024Quasiperiodic}. Such boundary sensitivity makes it necessary to examine the localization properties under OBC and PBC.

Quasiperiodic systems~\cite{Kohmoto1983MetalInsulator,Thouless1988Localization} provide a deterministic setting for studying localization beyond both periodic crystals\cite{Bloch1929Quantenmechanik, kittel2018introduction} and random disordered lattices\cite{Anderson1958Absence,THOULESS1974Electrons,Evers2008Anderson}. Owing to the absence of translational symmetry~\cite{Hatsugai1990Energy, jaric2012introduction}, their eigenstates can exhibit rich spatial structures, ranging from extended and localized states to critical states with multifractal fluctuations~\cite{Kohmoto1987Critical, Kohmoto1987Localization, Vladimir2009Optical, Deguchi2012QuantumCritical}. These features~\cite{Yao2019Critical,Wang2020Realization,Wang2021ManyBody, Xiao2021Observation,Gon2023Critical,Yang2024Exploring,Duncan2024Critical,Yao2024Wave,Zhang2025Critical} make quasiperiodic systems a useful setting for examining how non-Hermiticity modifies localization~\cite{Jiang2019Interplay,Tang2021Localization, Lin2022Topological,Junmo2023Localization,Zhou2023Non-Abelian, Shi2024Delocalization,Li2024Ring,Rangi2024Engineering, Wang2025Quasiperiodicity,Zheng2025Emergent, Li2025Anderson-skin, Gandhi2025Superconducting,liang2025size,Cai2026Quasicrystals,Zeng2026Coexistence}. When quasiperiodic modulation is combined with nonreciprocal hopping, an important question is whether quasiperiodic critical states can develop NHSE under OBC.

In this work, we consider an off-diagonal quasiperiodic Hatano--Nelson (HN) model~\cite{Hatano1996Localization, Hatano1997Vortex, Hatano1998Non-Hermitian}, whose open-boundary Hamiltonian can be transformed into an equivalent Hermitian quasiperiodic form by a non-unitary gauge transformation~\cite{Midya_Topological2024,Longhi_Erratic2025}. This mapping enables an exact analytical evaluation of the Lyapunov exponent in the thermodynamic limit and yields an expression for the phase boundary separating left-skin and right-skin regimes. By combining the analysis under OBC and PBC, we show that parameter regimes hosting quasiperiodic critical states under PBC can exhibit the NHSE when open boundaries are imposed. This comparison highlights the boundary-sensitive nature of critical localization in the presence of quasiperiodically modulated nonreciprocity. We further show that this behavior is not restricted to the nearest-neighbor single-band HN chain. Extending the analysis to long-range hopping models and to a two-band non-Hermitian Rice--Mele lattice with quasiperiodic modulations, we find that parameter regimes hosting quasiperiodic critical states under PBC can similarly exhibit the NHSE under OBC.

\section{Off-diagonal Hatano-Nelson model}
To study the interplay between quasiperiodic modulation and nonreciprocal hopping, we consider a quasiperiodically modulated HN chain whose Hamiltonian is given by
\begin{equation}
\hat{H}=\sum_{n=1}^{N-1}\left(J_{n}^{R} \hat{c}_{n+1}^{\dagger} \hat{c}_{n}+J_{n}^{L} \hat{c}_{n}^{\dagger} \hat{c}_{n+1}\right)+\hat{H}_{B}, 
\label{HN_model}
\end{equation}
where $N$ denotes the system size, $\hat{c}_{n}^{\dagger}$ ($\hat{c}_{n}$) creates (annihilates) a particle at lattice site $n$, and $J_{n}^{R}$ ($J_{n}^{L}$) represents the right (left) hopping amplitude. $\hat{H}_{B}$ is the Hamiltonian term that specifies the lattice boundary conditions: $\hat{H}_{B}=0$ corresponds to OBC, whereas for PBC, $\hat{H}_{B} = J_{N}^{R} \hat{c}_{1}^{\dagger} \hat{c}_{N} + J_{N}^{L} \hat{c}_{N}^{\dagger} \hat{c}_{1}$. Quasiperiodicity is introduced by specifying the site-dependent hopping amplitudes as: 
\begin{equation}
\begin{split}
{J}_{n}^{R}={t}_{{R}}+\mu_{{R}} \cos [2 \pi \alpha({n}+1 / 2)+\theta],\\
{J}_{n}^{L}={t}_{{L}}+\mu_{{L}} \cos [2 \pi \alpha({n}+1 / 2)+\theta].\\
\end{split}
\end{equation}
Here, $t_{R,L}$ denote the uniform components of the right and left hoppings, while $\mu_{R,L}$ set the strengths of the corresponding quasiperiodic modulations. The parameter $\theta$ is a global phase of the quasiperiodic modulation and does not affect the localization properties. The quasiperiodicity is governed by the irrational modulation frequency $\alpha$, chosen as the inverse golden mean $\alpha=(\sqrt{5}-1)/2$. In numerical calculations, this limit is realized by using a sequence of rational approximants $\alpha \simeq F_{m}/F_{m+1}$, where $F_m$ denotes the $m$th Fibonacci number. In the following, we focus on the modulated nonreciprocal case with $t_R=t_L$ and $\mu_R\neq\mu_L$, and examine how this quasiperiodic hopping imbalance gives rise to the NHSE under OBC.

\section{Non-Hermitian skin effect in quasiperiodic critical states}
\begin{figure}[t]
    \includegraphics[width=1\linewidth]{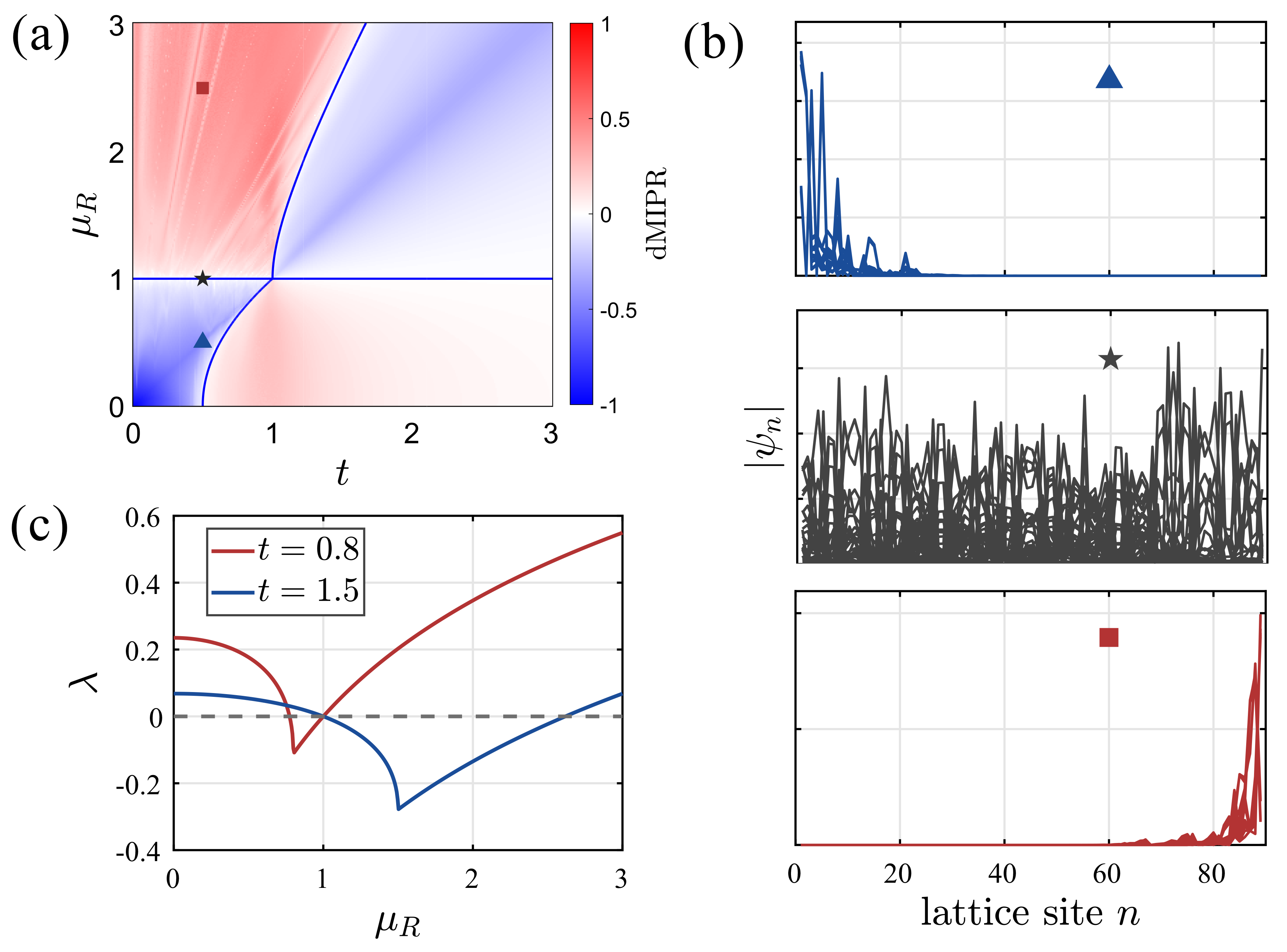}
    \caption{ 
    (a) The $\operatorname{dMIPR}$ under OBC as a function of $t$ and $\mu_{R}$ for $N=233$ and $\theta=0$. The blue line represents the theoretical prediction from Eq.~\eqref{critical_line}. The colored symbols mark the parameter points corresponding to the representative eigenstates shown in (b).  
    (b) Representative eigenstates under OBC at $t=0.5$ for $\mu_R=0.5$, $1$, and $2.5$ from top to bottom, corresponding to the marked points in (a).
    (c) The analytical $\lambda$ as a function of $\mu_{R}$. The red line corresponds to $t=0.8$, while the blue line represents $t=1.5$.} 
    \label{fig1}
\end{figure}

Under OBC, the single-particle eigenvalue equation of the model reads
\begin{equation}
E \psi_{n}=J_{n}^{L} \psi_{n+1}+J_{n-1}^{R} \psi_{n-1},
\label{the_spectral_problem}
\end{equation}
with $\psi_{0}=\psi_{N+1}=0$. The localization properties of the open HN chain can be analyzed using the non-unitary gauge transformation
\begin{equation}
\psi_n = \phi_n G_n,
\label{gauge_transformation}
\end{equation}
where $G_n=\prod_{l=1}^{n-1}\sqrt{J_{l}^{R}/J_{l}^{L}}$ and $G_1=1$.
This transformation corresponds to a local rescaling of the eigenstate amplitudes. Substituting the transformation into Eq.~\eqref{the_spectral_problem}, the eigenvalue equation is transformed into that of a quasiperiodic lattice with Hermitian hopping amplitudes 
\begin{equation}
E\phi_{n}=\sqrt{J_{n}^{R}J_{n}^{L}}\,\phi_{n+1}+\sqrt{J_{n-1}^{R}J_{n-1}^{L}}\,\phi_{n-1},
\label{Hermitian_equation}
\end{equation}
with $\phi_{0}=\phi_{N+1}=0$ under OBC.

To predict the phase boundary, we use the Lyapunov exponent $\lambda=\lim_{n\to\infty}(1/n)\ln|\psi_n/\psi_1|$~\cite{Hatano1996Localization,Avila2015Global}.
After introducing the non-unitary gauge transformation given by Eq.~\eqref{gauge_transformation}, we obtain:
\begin{equation}
\lambda =  \lim_{n \to \infty} \frac{1}{n} \ln \left| \frac{\phi_n}{\phi_1} \right| + \lim_{n \to \infty} \frac{1}{n}\ln \left|G_n\right| .
\label{lyapunov_decomposition}
\end{equation}
The first term in Eq.~\eqref{lyapunov_decomposition} represents the Lyapunov exponent associated with the equivalent reciprocal quasiperiodic lattice. In the parameter regimes considered here, the corresponding eigenstates are not exponentially localized. Therefore, this term vanishes in the thermodynamic limit. This means the $\lambda$ should be
\begin{equation}
\lambda =  \lim_{n \to \infty} \frac{1}{n}\ln \left|G_n\right| .
\end{equation}
In the thermodynamic limit ($n \to \infty$), it is found that
\begin{equation}
\begin{split}
\lambda & = \lim_{n \rightarrow \infty} \frac{1}{2n} \sum_{l=1}^{n-1} \ln \left|\frac{{t}_{{R}}+\mu_{{R}} \cos [2 \pi \alpha({l}+1 / 2)+\theta]}{{t}_{{L}}+\mu_{{L}} \cos [2 \pi \alpha({l}+1 / 2)+\theta]}\right| \\
& = \lim_{N \rightarrow \infty} \frac{1}{4\pi} \int_{0}^{2\pi} d\varphi 
    \ln \left|\frac{t_{R}+\mu_{R} \cos \varphi}{t_{L}+\mu_{L} \cos \varphi}\right|.
\end{split}
\end{equation}
Hence, we have
\begin{equation}
\lambda = \begin{cases}\dfrac{1}{2} \ln \dfrac{\mu_{R}}{\mu_{L}}, & \mu_{L} > t_{L} \text{ , } \mu_{R} > t_{R}, \\\\\dfrac{1}{2} \ln \dfrac{\mu_{R}}{t_{L} + \sqrt{t_{L}^{2} - \mu_{L}^{2}}}, & \mu_{L} \leqslant t_{L} \text{ , } \mu_{R} > t_{R}, \\\\\dfrac{1}{2} \ln \dfrac{t_{R} + \sqrt{t_{R}^{2} - \mu_{R}^{2}}}{\mu_{L}}, & \mu_{L} > t_{L} \text{ , } \mu_{R} \leqslant t_{R}, \\\\\dfrac{1}{2} \ln \dfrac{t_{R} + \sqrt{t_{R}^{2} - \mu_{R}^{2}}}{t_{L} + \sqrt{t_{L}^{2} - \mu_{L}^{2}}}, & \mu_{L} \leqslant t_{L} \text{ , } \mu_{R} \leqslant t_{R}.\end{cases}
\label{analytical_Lyapunov_exponent}
\end{equation} 
For the parameter regime considered below, we set $t_R=t_L=t$ and $\mu_L=1$.
The phase boundary in the thermodynamic limit is obtained from $\lambda=0$ in Eq.~\eqref{analytical_Lyapunov_exponent}. It consists of the following branches:
\begin{equation}
\mu_R =
\begin{cases}
1, & t<1, \quad (\mu_R>t),\\
t+\sqrt{t^2-1}, & t>1, \quad (\mu_R>t),\\
\sqrt{2t-1}, & 1/2 \leq t<1, \quad (\mu_R\leq t),\\
1, & t\geq 1, \quad (\mu_R\leq t).
\end{cases}
\label{critical_line}
\end{equation}

To verify this analytical boundary numerically and determine the direction of NHSE, we introduce the directional inverse participation ratio ($\operatorname{dIPR}$)~\cite{Wu2023Transition}:
\begin{equation}
\operatorname{dIPR}\left(\psi^{(j)}\right)=\mathcal{P}\left(\psi^{(j)}\right)\frac{\sum_{n=1}^{N} \lvert \psi_{n}^{(j)} \rvert^{4}}{\left(\sum_{n=1}^{N} \lvert \psi_{n}^{(j)} \rvert^{2}\right)^{2}}, 
\end{equation}
with $\mathcal{P}\left(\psi^{(j)}\right)$ defined as
\begin{equation} 
\mathcal{P}\left(\psi^{(j)}\right)=\operatorname{sgn}\left[\sum_{n=1}^{N}\left(n-\frac{N}{2}-\delta\right)\left|\psi_{n}^{(j)}\right|\right] .
\end{equation}
Here, $\delta$ is a positive value and is normally set to be $0 <\delta <0.5$. $\mathcal{P}\left(\psi^{(j)}\right)$ extracts the information of whether the eigenstate $\psi^{(j)}$ is more localized at the left or the right half of the lattice. The $\operatorname{dIPR}$ is positive for localization toward the right boundary and negative for localization toward the left boundary. We further present the phase diagram of the NHSE in Fig.~\ref{fig1}(a). The color bar corresponds to the mean values of the $\operatorname{dIPR}$ ($\operatorname{dMIPR}$) and is calculated as $\operatorname{dMIPR} = \frac{1}{N}\sum_{j=1}^{N} \operatorname{dIPR}\left(\psi^{(j)}\right)$. The blue line represents the theoretical phase boundary derived from Eq.~\eqref{critical_line} and shows excellent agreement with the numerical data.
To further illustrate the change of localization direction, Fig.~\ref{fig1}(b) shows representative eigenstates for fixed $t=0.5$ and different values of $\mu_R$. The eigenstates accumulate near the left boundary for small $\mu_R$, become delocalization near the transition point, and shift toward the right boundary as $\mu_R$ is further increased, consistent with the sign change of the $\operatorname{dMIPR}$ in Fig.~\ref{fig1}(a).

The Lyapunov exponent $\lambda$ captures the exponential decay rate of eigenstates, with $|\lambda|^{-1}$ corresponding to the localization length. The sign of $\lambda$ indicates the accumulation direction under OBC: $\lambda>0$ corresponds to accumulation toward the right boundary, whereas $\lambda<0$ corresponds to accumulation toward the left boundary. The $\lambda=0$ marks the point where the NHSE vanishes. The analytical $\lambda$ as a function of $\mu_{R}$ for fixed $t$ values, obtained from Eq.~\eqref{analytical_Lyapunov_exponent}, is shown in Fig.~\ref{fig1}(c). As $\mu_R$ is varied, $\lambda$ changes sign across the transition point, indicating a reversal in the direction of skin localization, while $\lambda=0$ is consistent with the phase boundary shown in Fig.~\ref{fig1}(a).

\begin{figure}[t]
    \includegraphics[width=1\linewidth]{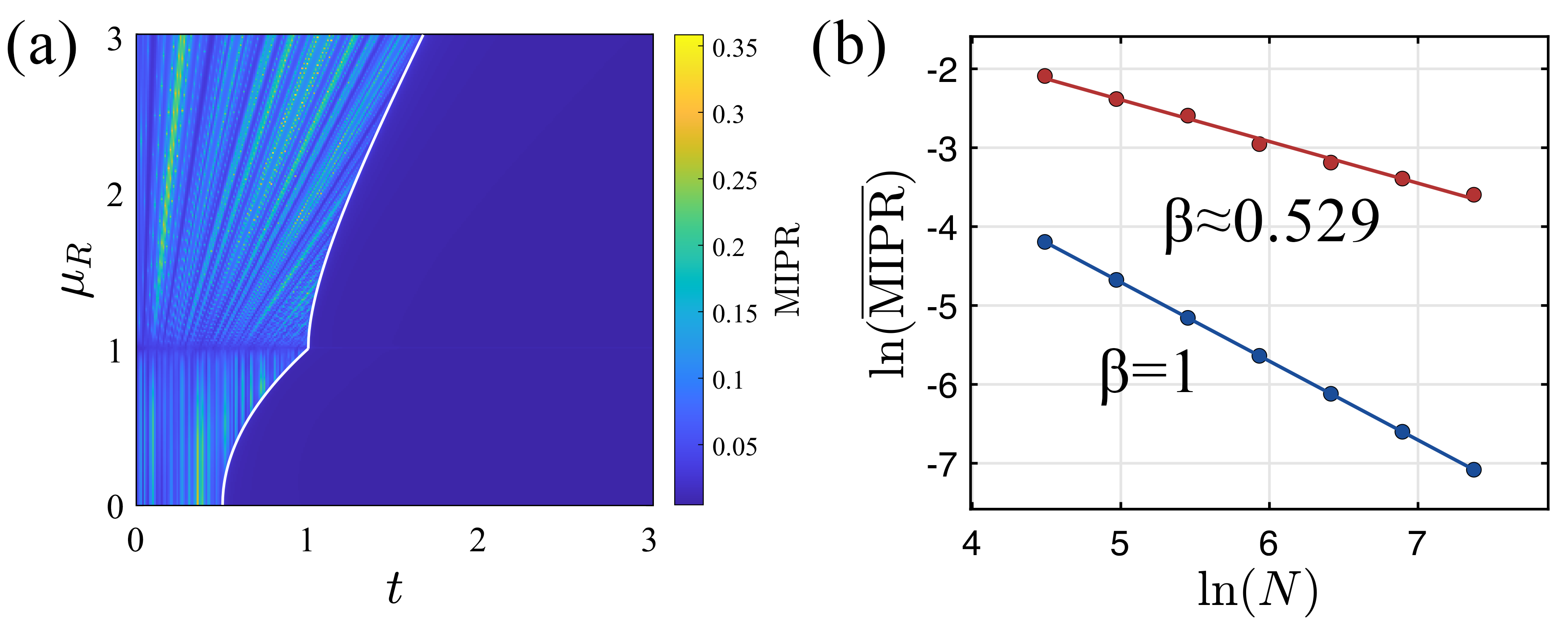}
    \caption{(a) The $\operatorname{MIPR}$ under PBC as a function of $t$ and $\mu_{R}$ for system size $N=233$ and $\theta=0$. The white line marks the phase transition line between the extended and critical regimes.
    (b) Finite-size scaling of the $\overline{\operatorname{MIPR}}$ versus system size $N$ on a logarithmic scale. The red points correspond to the parameter set with $t=0.3$ and $\mu_R=2.5$, while the blue points correspond to $t=2.5$ and $\mu_R=2$.
    The solid lines represent power-law fits. The statistical average is performed over $10^3$ random realizations of the global phase $\theta$. $\beta$ is the corresponding fractal dimension. } 
    \label{fig2}
\end{figure}

Having characterized the NHSE under OBC, we now analyze the eigenstates under PBC using the inverse participation ratio (IPR)~\cite{Bauer1990Correlation,Evers2000Fluctuations}. For a normalized eigenstate $\psi^{(j)}$, the IPR is defined as $\operatorname{IPR}^{(j)}=\sum_{n=1}^{N}|\psi_n^{(j)}|^4$. The mean inverse participation ratio ($\operatorname{MIPR}$) is defined as $\operatorname{MIPR} = \frac{1}{N}\sum_{j=1}^{N} \operatorname{IPR}^{(j)}$. 
Figure~\ref{fig2}(a) shows the $\operatorname{MIPR}$ as a function of $t$ and $\mu_R$ under PBC for $N=233$. The deep-blue region, where $\operatorname{MIPR}$ is close to zero, corresponds to extended eigenstates, whereas finite $\operatorname{MIPR}$ values indicate that the eigenstates are not extended. 

Since a finite MIPR at a fixed system size does not distinguish localized states from critical states, we further perform a finite-size scaling analysis of the MIPR~\cite{Hiramoto1989Scaling} under PBC. The corresponding fractal dimension $\beta$ is defined by $\beta=\lim _{N \rightarrow \infty}[\ln \operatorname{MIPR} / \ln (1 / N)]$. For localized states, $\beta \to 0$, and for extended states, $\beta \to 1$, whereas $0<\beta<1$ implies critical states. For a fixed value of lattice size $N$, the $\operatorname{MIPR}$ of the eigenstates is a random variable, which depends on the value of $\theta$. To extract the $\beta$, we perform a finite-size scaling analysis of $\overline{\operatorname{MIPR}} = \langle \operatorname{MIPR} \rangle_{\theta},$ defined as the average over realizations of the global phase $\theta$. Figure~\ref{fig2}(b) shows the finite-size scaling behavior of $\overline{\operatorname{MIPR}}$ for two representative parameter points. The red data exhibit a power-law decay with increasing system size, yielding a fractal dimension $\beta \approx 0.529$. This value indicates the critical nature of the corresponding eigenstates. The behavior of the $\overline{\operatorname{MIPR}}$ (blue points) in extended phase is also shown in Fig.~\ref{fig2}(b), exhibiting $\beta = 1$ as expected for extended states. Together with the OBC phase diagram in Fig.~\ref{fig1}(a), these results show that parameter regimes hosting quasiperiodic critical states under PBC can exhibit the NHSE under OBC.

\section{Generalization to long-range hopping Hamiltonian}
\begin{figure*}[t]
    \includegraphics[width=1\linewidth]{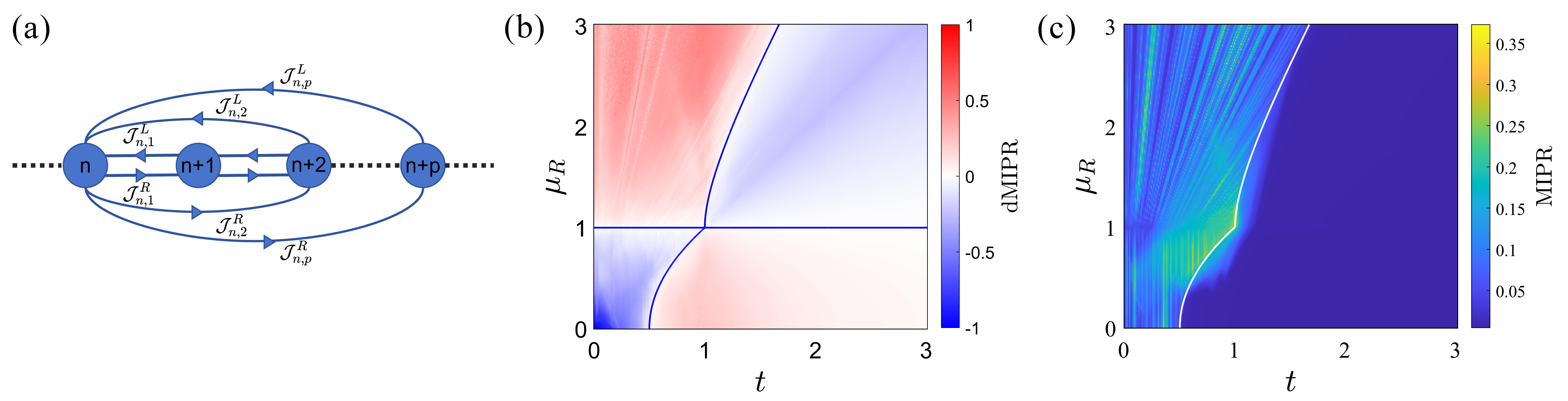}
    \caption{(a) Schematic illustration of the HN model with long-range hoppings.
    (b) The $\operatorname{dMIPR}$ under OBC as a function of $t$ and $\mu_{R}$.
    (c) The $\operatorname{MIPR}$ under PBC as a function of $t$ and $\mu_{R}$.
    Panels (b) and (c) correspond to a lattice with nearest and next-nearest neighbor hoppings, for $N=233$ and $\theta=0$.}
\label{fig3}
\end{figure*}

The preceding analytical mapping was derived for the nearest-neighbor model. To examine whether the NHSE persists beyond this minimal setting, we extend the HN model to include long-range hopping terms, as schematically illustrated in Fig.~\ref{fig3}(a). The generalized non-Hermitian Hamiltonian reads:
\begin{equation}
\hat{H}=\sum_{p=1}^{P}\big(\hat{H}^{(p)}+\hat{H}_B^{(p)}\big),
\end{equation}
where $\hat{H}^{(p)}$ denotes the $p$-th order hopping term and $\hat{H}_B^{(p)}$ is the corresponding boundary term under PBC. Here, $P$ is the maximum hopping range. For the class of long-range hoppings considered here, the $p$-th order right and left hopping amplitudes are defined as
\begin{equation}
\mathcal{J}^{R}_{n,p} = \prod_{i=0}^{p-1} J^{R}_{n+i},
\qquad
\mathcal{J}^{L}_{n,p} = \prod_{i=0}^{p-1} J^{L}_{n+i}.
\label{long_range_coupling}
\end{equation}
The corresponding hopping Hamiltonian is then given by
\begin{equation}
\hat{H}^{(p)} =\sum_{n=1}^{N-p}\left(\mathcal{J}^{R}_{n,p}\,\hat{c}_{n+p}^{\dagger}\hat{c}_{n}+\mathcal{J}^{L}_{n,p}\,\hat{c}_{n}^{\dagger}\hat{c}_{n+p}\right).
\label{long_range_Hp}
\end{equation}
For $p=1$, Eq.~\eqref{long_range_Hp} reduces to the nearest-neighbor Hamiltonian in Eq.~\eqref{HN_model}. For $p>1$, it describes hopping processes over longer distances, whose amplitudes inherit the quasiperiodic nonreciprocal modulation through Eq.~\eqref{long_range_coupling}.

Under OBC, the single-particle eigenvalue equation of the long-range model can be written as
\begin{equation}
E\psi_n=\sum_{p=1}^{P}\left(\mathcal{J}^{R}_{n-p,p}\psi_{n-p}+\mathcal{J}^{L}_{n,p}\psi_{n+p}\right).
\end{equation}
Under the non-unitary gauge transformation Eq.\eqref{gauge_transformation}, the single-particle eigenvalue equation reduces to
\begin{equation}
E\phi_n=\sum_{p=1}^{P}\left(\widetilde{\mathcal{J}}_{n-p,p}\phi_{n-p}+\widetilde{\mathcal{J}}_{n,p}\phi_{n+p}\right),
\end{equation}
where $\widetilde{\mathcal{J}}_{n,p}=\sqrt{\mathcal{J}^{R}_{n,p}\mathcal{J}^{L}_{n,p}}$.
The above mapping demonstrates that the long-range non-Hermitian Hamiltonian can be transformed into a Hermitian quasiperiodic Hamiltonian with effective hopping amplitudes $\widetilde{\mathcal{J}}_{n,p}$. The non-unitary gauge transformation completely removes the nonreciprocity even in the presence of long-range couplings. As a consequence, the condition $\lambda=0$ determining the left-right skin boundary, is still governed by the accumulated logarithmic imbalance between right and left hoppings. The detailed magnitude of the dMIPR, however, can differ from the nearest-neighbor case because long-range hopping modifies the finite-size eigenstate distributions.

Figure~\ref{fig3}(b) shows the $\operatorname{dMIPR}$ under OBC for a lattice with nearest- and next-nearest-neighbor hoppings. The phase structure is consistent with the analytical prediction.
For comparison, the $\operatorname{MIPR}$ under PBC is shown in Fig.~\ref{fig3}(c). The PBC phase diagram retains extended and critical regimes similar to those in Fig.~\ref{fig2}(a), indicating that the long-range hopping terms do not eliminate the quasiperiodic critical eigenstate structure. Combining the OBC and PBC results, we find that parameter regimes hosting critical eigenstates under PBC can also exhibit the NHSE under OBC in the long-range model.

\section{extension to multiband systems}
\begin{figure}[t]
    \includegraphics[width=1\linewidth]{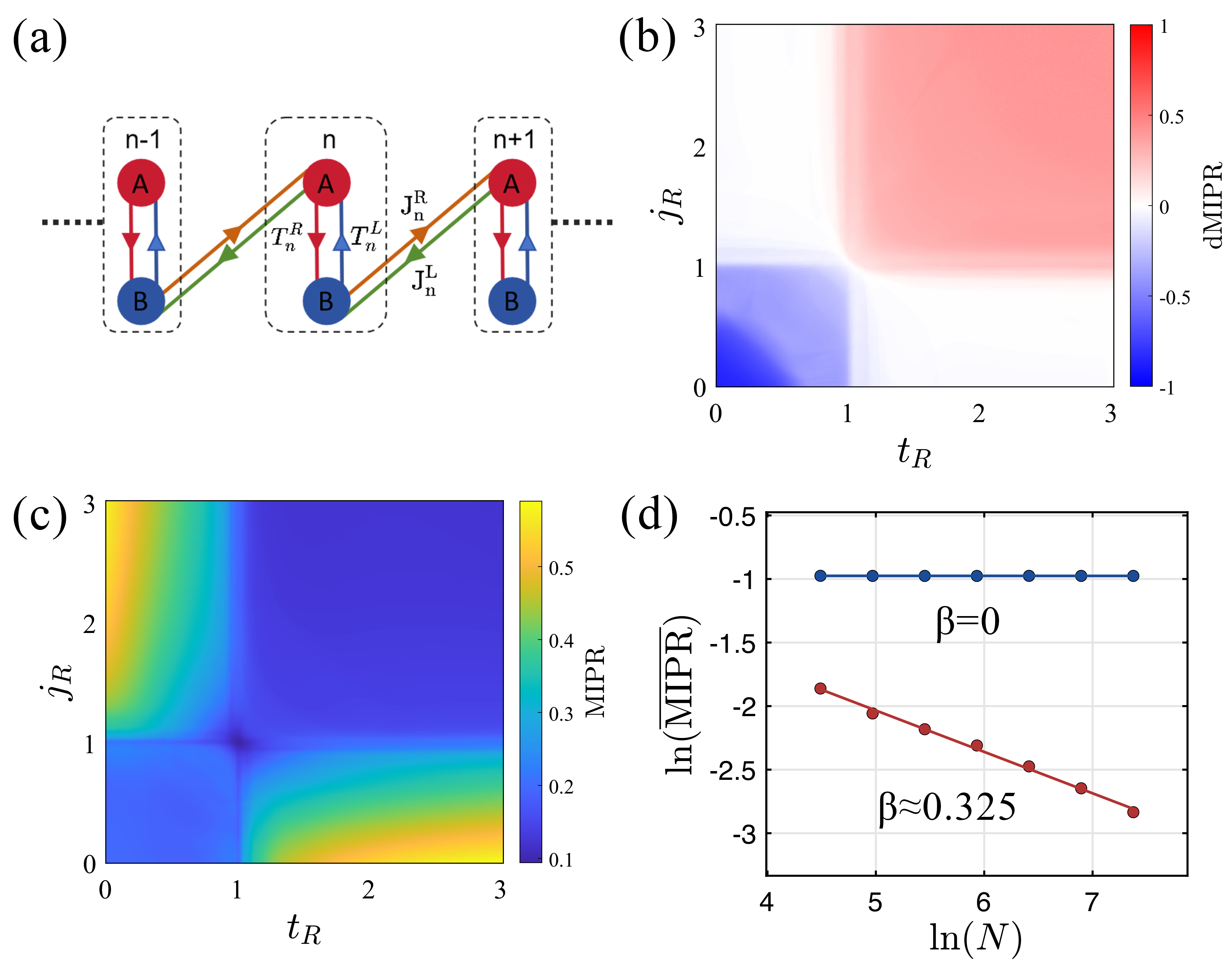}
    \caption{(a) Schematic of the Rice-Mele model with nonreciprocal hopping. 
    (b) The $\operatorname{dMIPR}$ under OBC as a function of $j_{R}$ and $t_{R}$ for $N=89$, $\Delta=1$, and $\theta=0$.
    (c) The $\operatorname{MIPR}$ under PBC as a function of $j_{R}$ and $t_{R}$ for $N=89$, $\Delta=1$, and $\theta=0$.
    (d) Finite-size scaling of the $\overline{\operatorname{MIPR}}$ versus system size $N$ on a logarithmic scale. The red points are obtained for $t_R=2.5$ and $j_R=3$, while the blue points are obtained for $t_R=0.5$ and $j_R=2$. The solid lines represent power-law fits, and the statistical average is performed over $10^3$ realizations of the global phase $\theta$.}
\label{fig4}
\end{figure}

The NHSE of quasiperiodic critical states under OBC can also appear beyond the single-band HN model. As a representative multiband extension, we consider a two-band non-Hermitian Rice--Mele model with quasiperiodic modulations, as schematically illustrated in Fig.~\ref{fig4}(a). The Hamiltonian of the system reads
\begin{equation}
\begin{aligned}
\hat{H} = &\sum_{n=1}^{N}\left(T_{n}^{L} \hat{c}_{n,A}^{\dagger}\hat{c}_{n,B}+T_{n}^{R}\hat{c}_{n,B}^{\dagger}\hat{c}_{n,A}\right)\\
&+\sum_{n=1}^{N-1}\left(J_{n}^{L}\hat{c}_{n,B}^{\dagger}\hat{c}_{n+1,A}+J_{n}^{R}\hat{c}_{n+1,A}^{\dagger}\hat{c}_{n,B}\right)\\
&+\sum_{n=1}^{N}\Delta\left(\hat{c}_{n,A}^{\dagger}\hat{c}_{n,A}-\hat{c}_{n,B}^{\dagger}\hat{c}_{n,B}\right)+ \hat{H}_{B},
\end{aligned}
\label{Rice-Mele_model}
\end{equation}
where $N$ is the number of unit cells, and $\Delta$ denotes the energy offset between the two sublattices. For the hopping amplitudes, we assume 
\begin{equation}
\begin{aligned}
T_{n}^{R} & =t_{{R}} \cos [2 \pi \alpha({n}+1 / 2)+\theta], \\T_{n}^{L}  &=t_{{L}} \cos [2 \pi \alpha({n}+1 / 2)+\theta], \\J_{n}^{R} & =j_{{R}} \cos [2 \pi \alpha({n}+1 / 2)+\theta], \\J_{n}^{L} &=j_{{L}} \cos [2 \pi \alpha({n}+1 / 2)+\theta].
\end{aligned}
\end{equation}
To reduce the number of independent parameters, we take the left hopping amplitudes as the reference scale and set $t_L=1$ and $j_L=1$.

Figure~\ref{fig4}(b) shows the $\operatorname{dMIPR}$ under OBC as a function of $t_R$ and $j_R$. The positive and negative regions in the phase diagram indicate that eigenstates can accumulate toward different boundaries in different parameter regimes.
We next compare the OBC results with the eigenstate properties under PBC.  Figure~\ref{fig4}(c) shows the $\operatorname{MIPR}$ as a function of $j_R$ and $t_R$ under PBC, where two distinct regimes can be observed. To determine the nature of these PBC eigenstates, we perform a finite-size scaling analysis of the $\overline{\operatorname{MIPR}}$ in Fig.~\ref{fig4}(d). For $j_R=3$ and $t_R=2.5$, $\overline{\operatorname{MIPR}}$ decreases with increasing system size, yielding a fractal dimension $\beta\approx0.325$. This indicates that the corresponding PBC eigenstates are critical. For $j_R=2$ and $t_R=0.5$, the $\overline{\operatorname{MIPR}}$ does not vanish as $N$ is increased and reaches a stationary value, corresponding to a fractal dimension $\beta=0$ and localized eigenstates. Combining the OBC and PBC results, we find that parameter regimes hosting quasiperiodic critical states under PBC can also exhibit the NHSE under OBC in this two-band model.

\section{conclusion}
In summary, we have analyzed a quasiperiodically modulated off-diagonal HN model in which nonreciprocity originates from unequal modulation strengths of the right and left hoppings. Under OBC, the non-unitary gauge transformation relates the eigenvalue equation to that of an effective Hermitian quasiperiodic lattice, allowing the Lyapunov exponent to be obtained analytically. This analytical result determines the phase boundary between the left-skin and right-skin regimes under OBC. By comparing the localization properties under PBC and OBC, we further clarify the boundary dependence of quasiperiodic critical states. Parameter regimes hosting quasiperiodic critical eigenstates under PBC can exhibit the NHSE under OBC. We have also examined representative extensions with long-range hopping and a two-band non-Hermitian Rice--Mele lattice. The results show that the appearance of the NHSE in quasiperiodic critical regimes is not limited to the nearest-neighbor single-band HN chain. These findings offer a concise perspective on the emergence of the NHSE in critical states of quasiperiodic systems and may provide guidance for future experimental realizations~\cite{Gong2018Topological,Zhang2019Non-Hermitian,Ozawa2019Topological,Li2020Topological,Gao2021NHHigherOrder,Hu2023Anti, Liu2024Complete,Huang2024AcousticResonances,Zhang2026Harmonic}.

\section*{Acknowledgments}
This work was supported by Natural Science Foundation of Jiangsu Province (Grant No. BK20231320).

\section*{DATA AVAILABILITY}
The data that support the findings of this article are not publicly available. The data are available from the authors upon reasonable request.

\bibliography{reference}
\end{document}